\documentclass[12pt]{article}
\usepackage{latexsym}
\begin{document}
\title{On the particularities of Bose-Einstein condensation of quasiparticles}{}
\author{
\textit
 {A. I. Bugrij${}^{a)}$, V. M. Loktev${}^{b)}$}\\
{\small N. N. Bogolyubov Institute of Theoretical Physics}\\ {\small of
the National Academy of Sciences of Ukraine,}\\ {\small ul.
Metrologicheskaya 14-b, Kiev 03680 GSR Ukraine} }
 \date{}
 \maketitle
 \begin{abstract}
 An attempt is made to determine the difference between Bose-Einstein
condensation of particles and quasiparticles. An equation is obtained for
the number of particles in a Bose-Einstein condensate as a function of the
total number of particles in the system. This equation is also written for
quasiparticles taking account of their creation by external pumping and
the presence of equilibrium thermal excitations in the system. Analyzing
both equations, the chemical potential of the pumped quasiparticles and
their number in the condensate are found as a function of the pumping
intensity. A condition under which the Bose-Einstein condensation of
low-energy quasiparticle excitations starts and occurs at any, including
quite high, temperatures is found.
\end{abstract}

1. It is well known that the Bose-Einstein condensation (BEC) is one of
the remarkable macroscopic quantum phenomena which can occur in one or
another collective of particles or quasiparticles (QP) with integer spin.
There are now a number of examples of the more or less direct experimental
observation of BEC, among which we call attention to BEC in rarified gases
of Bose atoms (Refs. 1 and 2; see also the review Ref. 3). Its critical
temperature is extremely low ($\sim $10$^{-6}$ K) because of the low
density of the particles and the relatively large mass of even the
lightest atoms. In this sense, quasiparticles are more interesting. Quite
high quasiparticle densities are experimentally fully achievable and the
effective masses, as a rule, are of the order the electronic masses (to
say nothing of the existence of massless qausiparticles$^{4})$. Possibly,
tor this reason the question of the BEC of QP for the example of
large-radius excitons (and biexcitons) was raised about 50 years
ago,$^{5}$ and the investigation of the their condensation became very
popular especially for indirect excitons or in heterostructures.$^{6}$ The
recent experiments on BEC of excitonic polaritons in microcavities also
merit equal attention$^{7,8}$. Here the condensation temperature, though
it increases by an order of magnitude, is still quite low---about several
degrees or fractions of a degree.

As concerns the short-radius dipole-active excitons, their condensation
has turned out to be practically unobservabie because of their short
lifetime relative to radiative decay and therefore the impossibility of
attaining the densities required for BEC. The high-density limitation
substantially weakens for magnons, which also can be regarded as a type of
Frenkel excitons (or, which is the same thing, short-radius excitons),
which possess a long lifetime. Apparently, it is this fact that helps to
achieve such high magnon densities, so that their condensation, as
asserted in Ref. 10, is observed in perfect yttrium-iron garnet films with
temperatures which are anomalously high and compared with room (!)
temperature. The works mentioned above and' the works of Demokritov's
group$^{11,12}$ devoted to BEC magnons make it necessary to analyze
whether or not the conditions at which BEC of QP becomes possible at truly
high temperatures are indeed satisfied, since nothing like this can be
achieved for particles and excitons.

2. First, we note a remarkable feature of BEC, which, being essentially a
phase transition, can occur in an ideal gas, which makes it possible to
calculate many physical charac-teristics and observable quantities of this
process to a high degree of accuracy. In addition, although objects which
do not interact with one another at all are unlikely to exist, the concept
of ideality could be even more characteristic for QP, which interact quite
weakly even at relatively high densities.

Nonetheless, it is not ideality (or weak ideality) that lies at the basis
of the main difference between particles and QP of the Bose type. The
fundamental difference between them is that if at a given density the
particles are characterized by a finite value of the chemical potential
$\mu $, then $\mu =0$ for equilibrium QP, since their density itself is
determined by only the temperature. This means that the realization and
observation of BEC of QP requires converting the system (for example, by
special nonthermal action) into an excited (in other words, a
nonequilibrium) state, which remains in existence for a sufficiently long
time (in any case, appreciably greater than the equilibration time and the
QP lifetime). Then the (quasi)equilibrium distribution of QP which is
established and which corresponds to BEC becomes accessible for
experimental study. But, on the other hand, it also follows from this that
one must take great care when describing the BEC of QP because of the
problematic nature of using the thermodynamic approach and the
nonstrictness of the application of various limits.

We note that there have been more than a few attempts (see, for example,
Refs. 13-15) at describing the kinetic behavior of the nonequilibrium QP
collective (specifically, ferromagnons\textbf{) }under one or another form
of pumping---pulsed, noise, and so forth, which can then condense. In such
strict approaches, additional information must be taken into account: on
the interaction of QP with one another, their interaction with other
objects --- phonons or defects, the character of the pumping, the temporal
evolution, which makes the model and the calculations quite complex. All
this, being undoubtedly important, nonetheless does not appear to be
determining for BEC itself. In complete correspondence with Bose and
Einstein, we understand it as a phenomenon of temperature redistribution
of a prescribed number of particles over the states and the accumulation
of a substantial fraction of them (particles) in the lowest of them
(states). Even if we confine ourselves to the assumption that pumping has
created only QP and that fast relaxation made them essentially equilibrium
states, we can raise many questions which follow from
experiments.$^{10-12}$ Neither is our objective to examine the coherent
manifestations of the QP collective accumulated in this manner in the
lowest (excited) state, which, certainly, is an interesting and topical
problem. We shall study the particularities which make it possible for QP
to condense at such high temperatures, including, in this case, room
temperatures; this problem is limited and has not yet been thoroughly
analyzed.

3. It is well known that the long-wavelength elementary quasiparticle
excitations determine the low-energy part of the spectrum of any
particular system. On the other hand, their interaction with one another
is the weakest interaction. As mentioned above, this makes it possible to
consider, to a first approximation, the subsystem of QP to be an ideal
gas. Let $\varepsilon ({\rm {\bf k}})=\varepsilon _0 +[\varepsilon ({\rm
{\bf k}})-\varepsilon _0 ]\equiv \varepsilon _0 +\varepsilon _{kin} ({\rm
{\bf k}})$ be the QP energy, and let $\varepsilon _0=\varepsilon ({\rm
{\bf k}}_{0})$ correspond to the lowest excited state. Then, according to
the assumption that the number of QP is conserved as a result of their
external interaction or, in other words, the fact that they have a finite
chemical potential, we write the number of QP in the \textbf{k}th quantum
state at fixed temperature as
\begin{equation}
\label{eq1} n({\rm {\bf k}},\mu )=\frac{1}{\exp\{[\varepsilon({\rm {\bf
k}})-\mu] /k_B T\}-1}\,.
\end{equation}
Instead of the quantity $\mu $, we introduce, following Ref. 16, a different
thermodynamic variable
\begin{equation}
\label{eq2}
 n_0 =\frac{1}{[\exp(\varepsilon _0 -\mu )/k_B T]-1},
\end{equation}
whence $\mu =\varepsilon _0 -k_B T\ln [(n_0 +1)/n_0 ]$. Using the expression
(\ref{eq2}). the occupation numbers (\ref{eq1}) can be easily rewritten in the form
\begin{equation}
\label{eq3} n({\rm {\bf k}},\mu )\to n({\rm {\bf k}},n_0
)=\frac{1}{(1+1/n_0 )\exp[\varepsilon _{kin} ({\rm {\bf k}}) /k_B
T]-1}=n_{\rm {\bf k}} (T)\frac{1}{1+[n_{\rm {\bf k}} (T)+1]/n_0 },
\end{equation}
where we have introduced the following notation for the number of particles
\begin{equation}
\label{eq4} n_{\rm {\bf k}} (T)=\frac{1}{\exp[\varepsilon _{kin} ({\rm
{\bf k}})/k_B T]-1},
\end{equation}
which, as one can see from its definition, does not depend on the gap in
the QP spectrum. The replacement (\ref{eq2}), eliminating the QP chemical
potential from the analysis, facilitates the study of the most interesting
and important situation where the nonphysical asymptotic limit $n_0 \to
\infty $ corresponds to the generally accepted condition for BEC $\mu \to
\varepsilon _0.^{1)}$

For sufficiently large values of $n_0 (\gg n_{\rm {\bf k}} (T))$, which,
generally speaking, should correspond to the BEC regime, the following
expansion becomes valid:
\begin{equation}
\label{eq5}
n_{\rm {\bf k}} (T,n_0 )=n_{\rm {\bf k}} (T)\left\{ {1-\left( {\frac{n_{\rm
{\bf k}} (T)+1}{n_0 }} \right)+\left( {\frac{n_{\rm {\bf k}} (T)+1}{n_0 }}
\right)^2-\cdot \cdot \cdot } \right\},
\end{equation}
which is a direct indication that the quantity $n_{\rm {\bf k}} (T)$ introduced
above is simply the maximum possible (corresponding to the formal condition
$n_0 \to \infty )$ occupation number of the corresponding state for fixed
$T.$

It is convenient to use the expansion (\ref{eq5}) when writing thermodynamically
equilibrium quantities. Specifically, the total (average) number of
particles becomes
\begin{equation}
\label{eq6}
N=\sum\limits_{\rm {\bf k}} {g_{\rm {\bf k}} n({\rm {\bf k}},n_0 )=} n_0
+\sum\limits_{{\rm {\bf k}}\ne {\rm {\bf k}}_0 } {g_{\rm {\bf k}} n({\rm
{\bf k}},n_0 )} \equiv n_0 +N_{exc} (T,n_0 ),
\end{equation}
where $g_{\rm {\bf k}} $ is the degeneracy of the state with wave vector
${\rm {\bf k}}$, and
\begin{eqnarray}
\nonumber N_{exc} (T,n_0 )&=&\sum\limits_{{\rm {\bf k}}\ne {\rm {\bf k}}_0
} {g_{\rm {\bf k}} n_{\rm {\bf k}} (T)\left\{ {1-\left( {\frac{n_{\rm {\bf
k}} (T)+1}{n_0 }} \right)+\left( {\frac{n_{\rm {\bf k}} (T)+1}{n_0 }}
\right)^2-\cdot \cdot \cdot } \right\}} \approx\\
\label{eq7}&\approx & N_{exc} (T)-\frac{\delta N_{exc} (T)}{n_0 },
\end{eqnarray}
corresponds to the total number of QP in the excited states. In the latter
equality, the following number is singled out:
\begin{equation}
\label{eq8} N_{exc} (T)=\sum\limits_{{\rm {\bf k}}\ne {\rm {\bf k}}_0 }
{g_{\rm {\bf k}} n_{\rm {\bf k}} (T)} ,
\end{equation}
which gives the maximum number (achieved under the same condition (see
above) $n_0 \to \infty )$ of all thermal excitations, and the coefficient
\begin{equation}
\label{eq9} \delta N_{exc} (T)=\sum\limits_{{\rm {\bf k}}\ne {\rm {\bf
k}}_0 } {g_{\rm {\bf k}} n_{\rm {\bf k}} (T)[n_{\rm {\bf k}} (T)+1]} ,
\end{equation}
is determined, to first order in $1/n_{0}$, by the fluctuations of the
quantity (\ref{eq7}). It is evident that irrespective of the specific form
of the QP dispersion law $\varepsilon _{kin} ({\rm {\bf k}})$ the
quantities $N_{exc} (T)$ and $\delta N_{exc} (T)$ are monotonically
increasing functions of the temperature, which is easily seen from the
definitions, since as $T$ increases, all occupation numbers increase.

Substituting (7) into (\ref{eq6}), taking account of (\ref{eq7}) and
(\ref{eq8}) and assuming $N\gg 1$ (and, correspondingly, $n_0 \gg 1)$, we
easily obtain the equation
\begin{equation}
\label{eq10} N\simeq n_0 +N_{exc} (T)-\frac{\delta N_{exc} (T)}{n_0 }
\end{equation}
for finding the number $n_{0}$ of Bose-condensed QP. We find from this
equation that the behavior of the desired number
\begin{equation}
\label{eq11} n_0 \equiv n_0 (T)=\frac{1}{2}\left\{ {N-N_{exc} (T)+\sqrt
{[N-N_{exc} (T)]^2+4\delta N_{exc} (T)} } \right\}
\end{equation}
for large $N$ and small $\delta N_{exc} (T)/N^2$ changes quite sharply when
the temperature crosses a certain value $T_c $ which is the solution of the
equation
\begin{equation}
\label{eq12} N_{exc} (T_c )=N,
\end{equation}
The Eq. (\ref{eq11}) shows directly that the critical temperature
introduced in this manner depends on the total number of particles in the
system and the specific form of the function $N_{exc} (T)$. The explicit
expression for the latter (as also for $T_c$)  is determined by the
spectrum, the dimensionality, and even the form of the system as well as
by the conditions at the boundaries of the system.

If we now introduce the dimensionless density $n_{BEC} (T)\equiv n_0
(T)/N$ of the Bose condensate, then in the limit $\delta N_{exc}
(T)/N^2\to 0$ it becomes a nonanalytic function of the temperature:
\begin{equation}
\label{eq13} n_{BEC} (T)=\left[ {1-\frac{N_{exc} (T)}{N}} \right]\theta
\left[ {1-\frac{N_{exc} (T)}{N}} \right],
\end{equation}
where $\theta (x)$ is a step function. Such behavior of the density
(\ref{eq12}) makes it possible to interpret it as the order parameter of
the BEC process in a given collective of particles or QP. We shall not
dwell on this, but we shall focus our attention on the question, raised
above, concerning the particularities which distinguish QP from particles.

4. Indeed, thus far the analysis could refer to both of these objects,
since the chemical potential was excluded from it. However, it is well
known and has already been mentioned that for QP $\mu =0$, to that any
system at finite $T$ will contain one or another --- thermally excited ---
quantity of them (compare Eq. (\ref{eq1})):
\begin{equation}
\label{eq14} n_{\rm {\bf k}}^{(th)} (T)=\frac{1}{\exp[\varepsilon ({\rm
{\bf k}})/k_B T]-1}.
\end{equation}
In complete analogy with Eq. (\ref{eq2}), it is also possible to single out the
equilibrium thermal occupation of the lowest state:
\begin{equation}
\label{eq15} n_0^{(th)} (T)\equiv n_0^{(th)} =\frac{1}{\exp(\varepsilon _0
/k_B T)-1},
\end{equation}
which is a given (increasing) function of the temperature.

Using the relation (\ref{eq14}), the numbers (\ref{eq13}) can be easily put into a form
analogous to (\ref{eq3}):
\begin{equation}
\label{eq16} n_{\rm {\bf k}}^{(th)} (T)\to n_{\rm {\bf k}}^{(th)}
(T,n_0^{(th)} )\equiv n_{\rm {\bf k}} (T)\frac{1}{1+\frac{n_{\rm {\bf k}}
(T)+1}{n_0^{(th)} }},
\end{equation}
where, remarkably, the notation $n_{\rm {\bf k}} (T)$ is identical to the
occupation number (\ref{eq4}) given above. Formally, the expression
(\ref{eq15}) is exact, but the number of thermal excitations, generally
speaking, does not have the quantitative advantages over the expression
used in the expansion (7) (at least, the leading terms) over the
occupation numbers $n_{\rm {\bf k}} (T)$ (or, especially, the numbers
$n_{\rm {\bf k}}^{(th)} (T))$ as $n_0 $ over $n_{\rm {\bf k}} (T)$ in the
BEC regime. However, it can be supposed that because the occupation
numbers of the excited states decrease exponentially, the error of the
series expansion in powers of $1/n_0^{(th)} $ will not be
substantial.$^{2)}$ Of course, for small $n_0^{(th)} $ the difference
between the particles and QP becomes less noticeable. However, if one
stays within the assumption that the initial temperatures are sufficiently
high (or energy of the elementary excitations is relatively low), then the
expansion can be written approximately as (compared Eq. (7))
\begin{equation}
\label{eq17} n_{\rm {\bf k}}^{(th)} (T,n_0^{(th)} )=n_{\rm {\bf k}}
(T)\left\{ {1-\left( {\frac{n_{\rm {\bf k}} (T)+1}{n_0^{(th)} }}
\right)+\left( {\frac{n_{\rm {\bf k}} (T)+1}{n_0^{(th)} }} \right)^2-\cdot
\cdot \cdot } \right\},
\end{equation}
and Eq. (\ref{eq17}) which follows immediately
\begin{equation}
\label{eq18} N_{th} \simeq n_0^{(th)} +N_{exc} (T)-\frac{\delta N_{exc}
(T)}{n_0^{(th)} },
\end{equation}
where, as mentioned, the quantities $N_{exc} (T)$ and $\delta N_{exc} (T)$
do not change, even if nonthermal (specifically, pumped) QP are present in
the system.

We now introduce their number according to the obvious expression
\begin{equation}
\label{eq19} n_{\rm {\bf k}} (T)=n_{\rm {\bf k}}^{(th)} (T)+n_{\rm {\bf
k}}^{(pump)} (T)
\end{equation}
for each ${\rm {\bf k}}$. Then, according to Eqs. (7) and (\ref{eq17}) it
is easily shown that for all ${\rm {\bf k}}\ne {\rm {\bf k}}_0 $
\begin{equation}
\label{eq20} n_{\rm {\bf k}}^{(pump)} (T)=n_{\rm {\bf k}} (T)[n_{\rm {\bf
k}} (T)+1]\left[ {\frac{1}{n_0^{(th)} (T)}-\frac{1}{n_0 (T)}}
\right]\left[ {1+\frac{n_{\rm {\bf k}} (T)+1}{n_0 (T)}} \right]^{-1}\left[
{1+\frac{n_{\rm {\bf k}} (T)+1}{n_0^{(th)} (T)}} \right]^{-1}.
\end{equation}
The latter expression simplifies substantially provided that the proposed
(and actually experimentally realizable$^{10-12})$ ratio of the parameters
of the system obtains: $k_B T\gg \varepsilon _0 ,\mu $. Indeed, we find
from Eq. (\ref{eq20}), taking account of the definition (\ref{eq2}), that
in this case the increment to the number of excitations created in all
states, with the exception of the lowest, by pumping can be represented in
the quite simple form
\begin{equation}
\label{eq21} n_{\rm {\bf k}}^{(pump)} (T)\approx \frac{\mu }{k_B T}n_{\rm
{\bf k}} (T)[n_{\rm {\bf k}} (T)+1].
\end{equation}
In other words pumping increases the initial occupation numbers somewhat,
but very little because of the ``temperature suppression factor'' that
appears $\mu /k_B T\ll 1$.

At the same time the density of the pumped QP in a Bose condensate always
behaves completely differently (see Eqs. (\ref{eq2}) and (\ref{eq15}):
\begin{equation}
\label{eq22} n_0^{(pump)} (T)=n_0 (T)-n_0^{(th)} (T)\approx \frac{k_B
T}{\varepsilon _0 }\frac{\mu }{(\varepsilon _0 -\mu )},
\end{equation}
or can be arbitrarily large because of the possibility $\mu \to
\varepsilon _0 $ (except for the exact equality $\mu =\varepsilon _0 )$.
It is remarkable that, conversely, in contrast to all $n_{\bf k}(T)$ with
${\rm {\bf k}}\ne {\rm {\bf k}}_0 $ the occupation numbers $n_0^{(pump)}
(T)$ increase additionally on account of (compare Eq. (\ref{eq21})) the
``temperature intensification factor'' $k_B T/\varepsilon _0 \gg 1.$

$5. $The role of pumping is obvious --- to create a definite number of QP.
But how is it related with the chemical potential? To answer this
question, we take account of the fact that (see Eq. (\ref{eq6})) the total
number $N$ of QP in the system can also be divided into two contributions:
the temperature-determined equilibrium part $N_{th}$ and the
pumping-determined number $N_{pump}$: $N=N_{th} +N_{pump} $. As is
well-known,$^{18}$ under electromagnetic pumping with intensity $I_{pump}$
this number changes in accordance with the simplest balance equation
$dN/dt=I_{pump} -(N-N_{th} )/\tau _{rel}$ where $\tau _{rel} $ is an
effective equilibration time after pumping is switched off (a more
accurate analysis is given in the Appendix). It follows immediately from
this equation that because of the obvious equality$^{3)} \quad dN_{th}
/dt=0$ actually becomes $dN_{pump} /dt=I_{pump} -N_{pump} /\tau _{rel} $
and that the stationary number (for times $t\gg \tau _{rel} $, when
$dN/dt=0$) is the total number of pumped particles, equal to
$N_{pump}^{st} =I_{pump} \tau _{rel} $. Assuming that the main number
(\ref{eq22}) of the latter accumulates precisely in the lowest state, or
that $n_0^{(pump)} (T)\approx N_{pump}^{st} $, we easily arrive at
\begin{equation}
\label{eq23} I_{pump} \tau _{rel} =\frac{k_B T}{\varepsilon _0 }\frac{\mu
}{(\varepsilon _0 -\mu )},
\end{equation}
which enables us to write the chemical potential of the pumped (and only
pumped) QP:
\begin{equation}
\label{eq24} \mu =\varepsilon _0 \frac{I_{pump} \tau _{rel} }{(k_B
T/\varepsilon _0 +I_{pump} \tau _{rel} )}.
\end{equation}
This expression shows that for relatively weak pumping, when $I_{pump}
\tau _{rel} \ll k_B T/\varepsilon _0 $, the chemical potential of the QP
is $\mu \approx \varepsilon_{0}^{2}I_{pump} \tau _{rel}/k_{B}T $, and for
strong pumping when $I_{pump} \tau _{rel} \gg k_B T/\varepsilon _0 (\gg
1)$ its value goes (see Eq. (\ref{eq24})) to its limit $\varepsilon _0 $,
always satisfying the physically necessary condition $\mu <\varepsilon _0
$. It is easy to show that the difference
\begin{equation}
\label{eq25} \varepsilon _0 -\mu \approx \frac{k_B T}{N_{pump}^{st} }\to
\frac{k_B T}{n_0^{(pump)} (T)}\approx \frac{k_B T}{I_{pump} \tau _{rel} },
\end{equation}
and, in consequence, as follows from Eqs. (\ref{eq23}) and (\ref{eq25}),
the thermodynamic properties of the collective of QP condensed as a result
of BEC should depend only on the ratio $I_{pump} \tau _{rel} /k_B T$. It
appears that such scaling behavior can be checked experimentally.

Finally, using Eqs. (7) and (\ref{eq18}), and subtracting the latter from
the former, gives
\begin{equation}
\label{eq26} n_0 -n_0^{(th)} +\frac{n_0 -n_0^{(th)} }{n_0 n_0^{(th)}
}\delta N_{exc} (T)=I_{pump} \tau _{rel} ,
\end{equation}
where the unknown is the number $n_0 (T)$ of QP in the condensate (or,
which is the same thing, the number $n_0^{(pump)} (T)$. Evidently, $n_0
(T)=n_0^{(th)} (T)$ in the absence of pumping. In addition, we find from
Eq. (\ref{eq26}) thai, specifically, for weak pumping, when $n_0^{(pump)}
(T)\ll n_0^{(th)} (T)$, the number of QP in the lowest level increases as
\begin{equation}
\label{eq27} n_0 (T)\approx n_0^{(th)} (T)+\left( {1-\frac{\varepsilon
_0^2 }{(k_B T)^2}\delta N_{exc} (T)} \right)I_{pump} \tau _{rel} .
\end{equation}
For strong pumping and when nonequilibrium has been reached $n_0^{(pump)}
(T)\gg n_0^{(th)} (T)$, the number of QP in the lowest level is given by
the relation
\[
n_0 (T)\approx n_0^{(pump)} (T)=I_{pump} \tau _{rel} -\frac{\varepsilon _0
}{k_B T}\delta N_{exc} (T).
\]
Although it is obvious that the latter number also increases as the pump
intensity, a ``transition" occurs from one straight line onto another
displaced below the origin of the coordinates, which also can be checked
experimentally.

However, if the small corrections to the linear behavior of the number of
QP in the Bose condensate as a function of the pumping intensity are
neglected, which is precisely what was assumed in the derivation of the
chemical potential, then we arrive at a seemingly paradoxical result: the
pumped low-frequency QP are always in the BEC regime. However, one must
remember that the so-called temperatures must satisfy the inequality
$k_{B}T\gg \varepsilon_{0}$.

This can be stated differently: the computed change of the occupation
numbers of the quasiparticle states (in the experiments Refs. 10-12 this
is a ferromagnetic plate exposed to external pumping) shows that the
initial---thermally equilibrium---QP ``prevent'' the accumulation of new
QP in all states except the lowest one (see Eqs. (\ref{eq21}) and
(\ref{eq22})). For this reason, it seems that pumping-created QP have
nothing to do but to collect exclusively in it. As follows from our
analysis, this will occur at all times (of the order of the action of the
pumping pulse) irrespective of the temperature as soon as, once again,
$k_{B}T\gg\varepsilon_{0}$. Thus, the satisfaction of this inequality
makes the temperature a factor which not only does not impede (neglecting
relaxation processes, which decrease $\tau_{rel}$)
 but rather even promotes BEC to some extent. Of
course, as mentioned above, the latter assertion relies on the assumption
that the temperature remains constant. In addition, temperature stability
in a real experiment requires special attention. However, it is worth
repeating that when high values of the temperature are reached, its
relative increase as a result of irradiation processes should not be too
large.

6. The results obtained above show that certain types of QP, specifically,
QP with low activation energy (or at temperatures not exceeding them)
under external pumping can (more accurately, must) accumulate only in the
lowest possible state, which corresponds completely with the BEC
phenomenon. In addition, such accumulation is accompanied by relaxation
from the strongly nonequilibrium state into (thermal) a
(quasi)equilibriurn state which is found to be indistinguishable from a
Bose-condensed state. It appears that experimental verification (including
by measuring the Mandel'shtam-Brillouin scattering$^{10-12,19})$  of the
computational results obtained in the present work could show that there
is a large difference between BEC particles (for example, Bose atoms of
light metals) and low-frequency QP. This difference could be manifested,
first and foremost, even in its (BEC) most important characteristic---the
critical temperature. Other features (specifically, coherent and
fluctuation properties) of BEC for QP of different nature, with different
dispersion laws, and in systems with different dimensionality will be
analyzed elsewhere.

One of us (A.I.B.) wishes to thank M. I. Gorenshtein for helpful and
stimulating discussions concerning BEC. We also thank S. A. Demokritov and
G.A. Melkov for their support and interest, associated with the desire to
organize experiments to verify this work, as well as for suggestions and
remarks. Finally, this work was supported in part by Project No. 10/07-N
of the target program "Nanosize systems, nano-materials, and
nanotechnologies" of the National Academy of Sciences of Ukraine.

\begin{center}
\textbf{APPENDIX}
\end{center}

To simplify the exposition the balance equation given above for
determining the number of pumped QP was written in a form which gives only
the proportionality between $N^{st}_{pump }$and the absorbed power
$I_{pump}\tau_{rel}$; there are no terms which give rise to a pump
threshold near which $N_{pump}\sim\sqrt{I_{pump}}.^{12,20}$

In this connection we recall that under the conditions of parametric
electromagnetic excitation the exponential growth of the number of magnons
in the system occurs when the creation rate of these QP is higher than
their annihilation rate, whose determining channel are pair
collisions.$^{21}$ Then, to take these circumstances into account we shall
write the balance equation in the form (see Ref. 19)
$$\frac{dN}{dt}=I_{pump}(N+1)-\frac{N-N_{\rm th}}{\tau_{rel}^{(1)}}-
\frac{(N-N_{\rm th})^{(2)}}{\tau_{rel}^{2}}\, ,$$ or, in accordance with
the notation adopted,
$$
\label{eq(A1)}
\frac{dN_{pump}}{dt}=\widetilde{I}_{pump}+\left(I_{pump}-\frac{1}{\tau_{rel}^{(1)}}\right)N_{pump}-\frac{N_{pump}^{2}}{\tau_{rel}^{(2)}}\,,\eqno{(A1)}
$$ where $\widetilde{I}_{pump}\equiv I_{pump}(N_{\rm th}+1)$, and
$\tau_{rel}^{(1)}$ and $\tau_{rel}^{(2)}$ are the lifetimes of the magnons
with respect to one- and two-particle processes.

It is easily shown from Eq. (Al) that the threshold behavior of the number
of pumped QP is completely determined by the sign of the second term on
the right-hand side of this equation, and the last term limits the growth
of their number. It is precisely the equality
$I_{pump}=1/\tau_{rel}^{(1)}$ that sets the magnitude of the desired
threshold, as happens in the $S$ theory of parametric resonance. $^{21}$

As for the stationary number of pumped magnons, according to the same
equation (Al) $N_{pump}^{\rm st}\sim\widetilde{I_{pump}}\tau_{rel}^{(1)}$
which obtained above; near threshold, when
$I_{pump}\tau_{rel}^{(1)}\sim1$, the behavior changes and $N_{pump}^{\rm
st}\sim\sqrt{\widetilde{I}_{pump}\tau_{rel}^{(2)}}$\,; finally, if
$I_{pump}\gg1/\tau_{rel}^{(1)}$, the number $N_{pump}^{\rm st}\approx
I_{pump}\tau_{rel}^{(2)}/2+\sqrt{(I_{pump}\tau_{rel}^{(2)}/2)^{2}+\widetilde{I}_{pump}\tau_{rel}^{(2)}}$,
i.e. it is described by a dependence which is, once again, close to
linear. This is the qualitative picture observed experimentally, where the
function $N_{pump}^{\rm st}(I_{pump})\approx n_{0}^{pump}(I_{pump})$
changes from linear to square-root back to linear behavior.$^{20}$ If
necessary, the corresponding dependences can be easily introduced into
the. relations (\ref{eq23})-(\ref{eq26}), whose meaning remains intact.

\bigskip
$^{a)}$E-mail: abugrij@bitp.kiev.ua

$^{b)}$E-mail: vloktek@bitp.kiev.ua

$^{1)}$We note that in the theoretical study in Ref. 17 of the
fluctuations of a Bose-condensate of ultracold atoms the chemical
potential was also eliminated from the expression for the number of
particles but in a different manner so that the noted divergence of the
quantity remained.

$^{2)}$Here we implicitly assume that $n_{0}^{\rm th}\gg1$, which can hold
only under the obvious condition $k_{B}T\gg \varepsilon_{0}$  and holds
with room to spare, for example, in experiments with BEC of
ferromagnets.$^{10-12}$ At the same time $n_{0}^{\rm
th}\sim\exp(-\varepsilon/k_{B}T)\ll1$ almost always holds for excitons or
QP with a large gap, whence it follows that the Boltzmann distribution
describes their thermal number. Here $n_{\rm k}^{\rm th}(T)\approx n_{\rm
k}(T)\exp(-\varepsilon_{0}/k_{B}T)\ll n_{\rm k}(T)$.

$^{3)}$Of course, neglecting a possible change of temperature, whose
relative magnitude near 10$^{2}$ K cannot be large.

\bigskip
{\bf REFERENCES}

$^{1}$ M. H. Anderson, J. R. Ensher, M. R. Matthews, C. E. Wieman, and E.
A. Cornell,

\ \ Science \textbf{269}, 198 (1995).

$^{2} $ K. B. Davis, M.-O. Mewes, M. R. Andrews, N. J. van Drateu, D. S.
Durfee,

\ \ D. M. Kurn, and W. Keterle, Phys. Rev. Lett. \textbf{75}, $3969
$(1995).

$^{3}$ F. Dalfovo, S. Giorgini, L. P. Pitaevskii, and S. Stringari, Rev.
Mod. Phys. \textbf{71},

\ \ 463 (1999).

$^{4}$N. B. Brandt and V. A. Kul'bachinskii, \textit{Quasiparticles in the
Physics of Condensed}

\ \ \textit{Media, }Fizmatlit, Moscow (2005).

$^{5}$S. A. Moskalenko, Fiz. Tverd. Tela (Leningrad) \textbf{4}, 276 (1962).

$^{6}$V. B. Timofeev, Usp. Fiz. Nauk \textbf{175}, 315 (2005).

$^{7}$J. Kasprzak, M. Richard, S. Kundermann, A. Baas, J. M. J. Keeling, F.
M.Marchetti,

\ \ M. H. Szymanska, R. Andre, J. L. Staehli, V. Savona, P. B.Littlewood,
B. Deveaud,

\ \ and Le Si Dang, Nature (London) \textbf{443}, 409 (2006).

$^{8}$S. D. Snoke. Nature (London) \textbf{443}, 403 (2006).

$^{9}$Yu. B. Gaididei, V. VI. Loktev, and A. E. Prikhot'ko, Fiz. Nizk.
Temp. \textbf{3}, 549 (1977)

\ \ [Sov. J. Low Temp. Phys. \textbf{3}. 263 (1977)].

$^{10}$S. O. Demokritov, V. E. Demidov, O. Dzyapko, G. A. Melkov, A.
A.Serga, B.Hillebrands,

\ \ and A. N.Slavin. Nature (London) \textbf{443}, 430 (2006).

$^{11}$V. E. Demidov, O. Dzyapko, S. O. Demokritov, G. A. Melkov, and A.
N.Slavin,

\ \ Phys. Rev. Lett. \textbf{99}, 037205 (2007).

$^{12}$V. E. Demidov, O. Dzyapko, S. O. Demokritov, G. A. Melkov, and A.
N.Slavin,

\ \ Phys. Rev. Lett. \textbf{100}, 047205 (2008).

$^{13}$Yu. D. Kalafati and V. L. Safonov, JETP Lett. \textbf{50}, 149
(1989).

$^{14}$G.A. Melkov, V. L. Safonov, A. Yu. Taranenko, and S. V. Sholom,

\ \ J.Magn. Magn. Mater. \textbf{132}, 180 (1994).

$^{15}$Yu. E. Lozovik, A. G. Semenov, and M. Willander, Pisma ZhETF
\textbf{84}, 176 (2006).

$^{16}$A. M. Bugrij and V. M. Loktev, Fiz. Nizk. Temp. \textbf{33}, 51
(2007)

\ \ [Low Temp. Phys. \textbf{33}, 37 (2007)].

$^{17}$H. D. Politzer, Phys. Rev. A \textbf{54}, 5048 (1996).

$^{18}$V. M. Fain and Ya. N. Khanin. \textit{Quantum Radiophysics, }Sov. Radio, Moscow (1965).

$^{19}$V M. Loktev, Fiz. Nizk. Temp. \textbf{34}, 231 (2008) [Low Temp.
Phys. \textbf{34}, 178 (2008)].

$^{20}S. $O. Demokritov, V. E. Demidov, O. Dzyapko, G. A. Melkov, and A.
N. Slavin,

\ \ New J. Phys. \textbf{10}, 045029 (2008).

$^{21}$V. E. Zakharov, V. S. L'vov. and S. S. Starobinets, Usp. Fiz. Nauk
\textbf{114}, 609 (1974).

\end{document}